\documentclass[pra,letterpaper,twocolumn,showpacs,floatfix,superscriptaddress,amsmath,preprintnumbers,citeautoscript,aps,longbibliography]{revtex4-1}
\pdfoutput=1
\usepackage[T1]{fontenc}\usepackage[latin1]{inputenc}
\usepackage{dcolumn,graphicx,color,booktabs,microtype,afterpage} \graphicspath{{Figures/}}
\usepackage[charter,greekuppercase=italicized]{mathdesign}
\renewcommand{\figurename}{Fig.}
\renewcommand{\tablename}{Table}
\makeatletter\renewcommand{\fnum@figure}[1]{\figurename~\thefigure.}\makeatother
\makeatletter\renewcommand{\fnum@table}[1]{\tablename~\thetable.}\makeatother
\newcount\hh \newcount\mm
\hh=\time \divide\hh by 60
\mm=\hh \multiply\mm by 60 \mm=-\mm
\advance\mm by \time
\def\now{\number\hh:\ifnum\mm<10{}0\fi\number\mm}
\usepackage[colorlinks,plainpages=false,linkcolor=blue,urlcolor=blue,citecolor=blue,pdfpagemode=UseNone,pdfstartview=FitBH]{hyperref}

\newcommand{\YbAl}{YbAlO$_3$}
\newcommand{\DySc}{DyScO$_3$}
\newcommand{\YbFe}{YbFeO$_3$}

\begin{document}

\title{Antiferromagnetic ordering and dipolar interactions of YbAlO$_3$}

\author{L.~S.~Wu}
\affiliation{Neutron Scattering Division, Oak Ridge National Laboratory, Oak Ridge, TN 37831, USA}
\affiliation{Department of Physics, Southern University of Science and Technology, Shenzhen 518055, China}

\author{S.~E.~Nikitin}
\affiliation{Max Planck Institute for Chemical Physics of Solids, N\"{o}thnitzer Str. 40, 01187 Dresden, Germany}
\affiliation{Institut f\"ur Festk\"orper- und Materialphysik, Technische Universit\"at Dresden, D-01069 Dresden, Germany}

\author{M.~Brando}
\affiliation{Max Planck Institute for Chemical Physics of Solids, N\"{o}thnitzer Str. 40, 01187 Dresden, Germany}

\author{L.~Vasylechko}
\affiliation{Lviv Polytechnic National University, 79013 Lviv, Ukraine}

\author{G.~Ehlers}
\affiliation{Neutron Technologies Division, Oak Ridge National Laboratory, Oak Ridge, TN 37831, USA}

\author{M.~Frontzek}
\affiliation{Neutron Scattering Division, Oak Ridge National Laboratory, Oak Ridge, TN 37831, USA}

\author{A.~T.~Savici}
\affiliation{Neutron Scattering Division, Oak Ridge National Laboratory, Oak Ridge, TN 37831, USA}

\author{G.~Sala}
\affiliation{Neutron Scattering Division, Oak Ridge National Laboratory, Oak Ridge, TN 37831, USA}

\author{A.~D.~Christianson}
\affiliation{Materials Science and Technology Division, Oak Ridge National Laboratory, Oak Ridge, Tennessee 37831, USA}
\affiliation{Neutron Scattering Division, Oak Ridge National Laboratory, Oak Ridge, TN 37831, USA}

\author{M.~D.~Lumsden}
\affiliation{Neutron Scattering Division, Oak Ridge National Laboratory, Oak Ridge, TN 37831, USA}

\author{A.~Podlesnyak}
\thanks{Corresponding author: podlesnyakaa@ornl.gov}
\affiliation{Neutron Scattering Division, Oak Ridge National Laboratory, Oak Ridge, TN 37831, USA}

\date{\today}

\begin{abstract}

In this paper we report low-temperature magnetic properties of the rare-earth perovskite material \YbAl.
Results of elastic and inelastic neutron scattering experiment, magnetization measurements along with the crystalline electrical field (CEF) calculations suggest that the ground state of Yb moments is a strongly anisotropic Kramers doublet, and the moments are confined in the $ab$-plane, pointing at an angle of  $\varphi = \pm 23.5^{\circ}$ to the $a$-axis.
With temperature decreasing below $T_{\rm N}=0.88$~K, Yb moments order into the coplanar, but non-collinear antiferromagnetic (AFM) structure $AxGy$, where the moments are pointed along their easy-axes.
In addition, we highlight the importance of the dipole-dipole interaction, which selects the type of magnetic ordering and may be crucial for understanding magnetic properties of other rare-earth orthorhombic perovskites.
Further analysis of the broad diffuse neutron scattering shows that one-dimensional interaction along the $c$-axis is dominant, and suggests \YbAl\ as a new member of one dimensional quantum magnets.
\end{abstract}

\maketitle

\section{Introduction}

Perovskite materials with chemical composition $RM$O$_3$ ($R$ is a trivalent rare-earth ion and $M$ is a $3d$ transition metal) are in a focus of attention in modern solid-state physics and materials science, because they exhibit interesting magnetic~\cite{white,plakhty1983Yb,Nikitin2018,Wu2019}, multiferroic~\cite{Cheong2007} and optical~\cite{kimel2004} effects.
A number of intriguing magnetic properties arise in these materials from the coupling between the $4f$ and $3d$ magnetic sublattices, but its accurate microscopic description is still absent.
On the one hand, it is well established from both experimental and theoretical sides, that the strong Heisenberg superexchange interaction $M - O - M$ induces a robust antiferromagnetic (AFM) order below $T_{\rm N}~\gg~100$~K~\cite{hahn2014,shapiro1974,yamaguchi1973symmetry,Yamaguchi1974}.
In contrast, the $R - M$ interaction is much weaker and  usually taken into account phenomenologically, while the $R - R$ interaction is often not even considered~\cite{bazaliy2005,Belov1976,Yamaguchi1974}.

Our recent research of YbFeO$_3$ revealed that Yb moments form spin chains along the $c$-axis and exhibit unconventional low-energy spin excitations, which are strongly modified by the presence of the magnetic Fe sublattice~\cite{Nikitin2018}.
Contrary to the almost isotropic Fe spins, the Yb moments exhibit a strong single-ion anisotropy, which is rather important for an understanding of the magnetic properties of this material~\cite{bazaliy2005}.
Results of this work raised two important questions: (i) What is the magnetic ground state and magnetic anisotropy of the Yb moments in a distorted perovskite lattice? (ii) How one can efficiently describe Fe-Yb interaction?
While the second question requires advanced theoretical DFT-based calculations, one can answer the first one by studying isostructural Yb$M$O$_3$ materials with a nonmagnetic ion in $M$ position.
To this end, we performed a comprehensive study of YbAlO$_3$, and demonstrated that it provides a realization of a quantum spin $S = 1/2$ chain material exhibiting both quantum critical Tomonaga-Luttinger liquid behavior and spinon confinement-deconfinement transitions in different regions of magnetic field-temperature  phase space~\cite{Wu2019}.

In this work we focus on the magnetic ground state properties of YbAlO$_3$ and discuss its single-ion anisotropy and moment configuration at low temperature.
We performed neutron scattering and magnetization measurements combined with point charge model \textit{ab-initio} calculations.
We put the results in context with the findings from a previous study of the iso-structural compound \DySc~\cite{Wu2017}.
We show that the combination of strong spin-orbit coupling and crystalline electrical field effects creates an energetically isolated Kramers doublet ground state in both systems.
The ground state doublets have a strong uniaxial anisotropy, which constrains the magnetic moments in the $ab$-plane with angle $\alpha$ to the $a$-axis, and depending on $\alpha$ dipolar interchain interaction determine the type of AFM ordering.
However, distinct from \DySc, in which case neither the transverse nor the longitudinal fluctuations are seen in inelastic neutron scattering, due to the strong Ising single-ion anisotropy and the constraint of the  $\Delta{S}=1$ selection rule~\cite{Wu2017}, the analysis of the \YbAl\ ground states wavefunctions shows that the longitudinal fluctuations are visible to neutrons, making it a perfect object for exploring one-dimensional quantum magnetism.

\section{Experimental Details}

A single crystalline sample of \YbAl\ with clear orange color was prepared by the Czochralski technique, as described elsewhere~\cite{Buryy2010,Noginov2001}.
Magnetization measurements were performed using a Quantum Designs Magnetic Property Measurement System (MPMS) with a horizontal sample rotator insert and MPMS-3 with $^3$He insert for the low-temperature measurements down to 0.5~K.
Neutron scattering measurements of \YbAl\ were performed at the time-of-flight Cold Neutron Chopper Spectrometer (CNCS)~\cite{CNCS1,CNCS2}, at the Spallation Neutron Source (SNS) at Oak Ridge National Laboratory.
Data were collected with a single crystal \YbAl\ sample of mass around 0.6~g, which was aligned in the $(0KL)$ scattering plane.
A bottom-loading dilution refrigerator insert was used to access temperatures as low as 50~mK.
The incoming neutron energy was fixed at 3.32~meV ($\lambda_{i}=4.97$~{\AA}) and 50~meV ($\lambda_{i}=1.28$~{\AA}), and the high-flux instrument mode was used to maximize the neutron intensity.
The software packages \textsc{Dave}~\cite{Dave} and \textsc{MantidPlot}~\cite{Mantid} were used for data reduction and analysis.
The crystal electric field (CEF) calculations were performed using the \textsc{McPhase} software package~\cite{McPhase}.

\section{Results and Analysis}

\subsection{Crystal structure and crystal electric field}

\begin{figure}
  \centering
  \includegraphics[width=0.7\columnwidth]{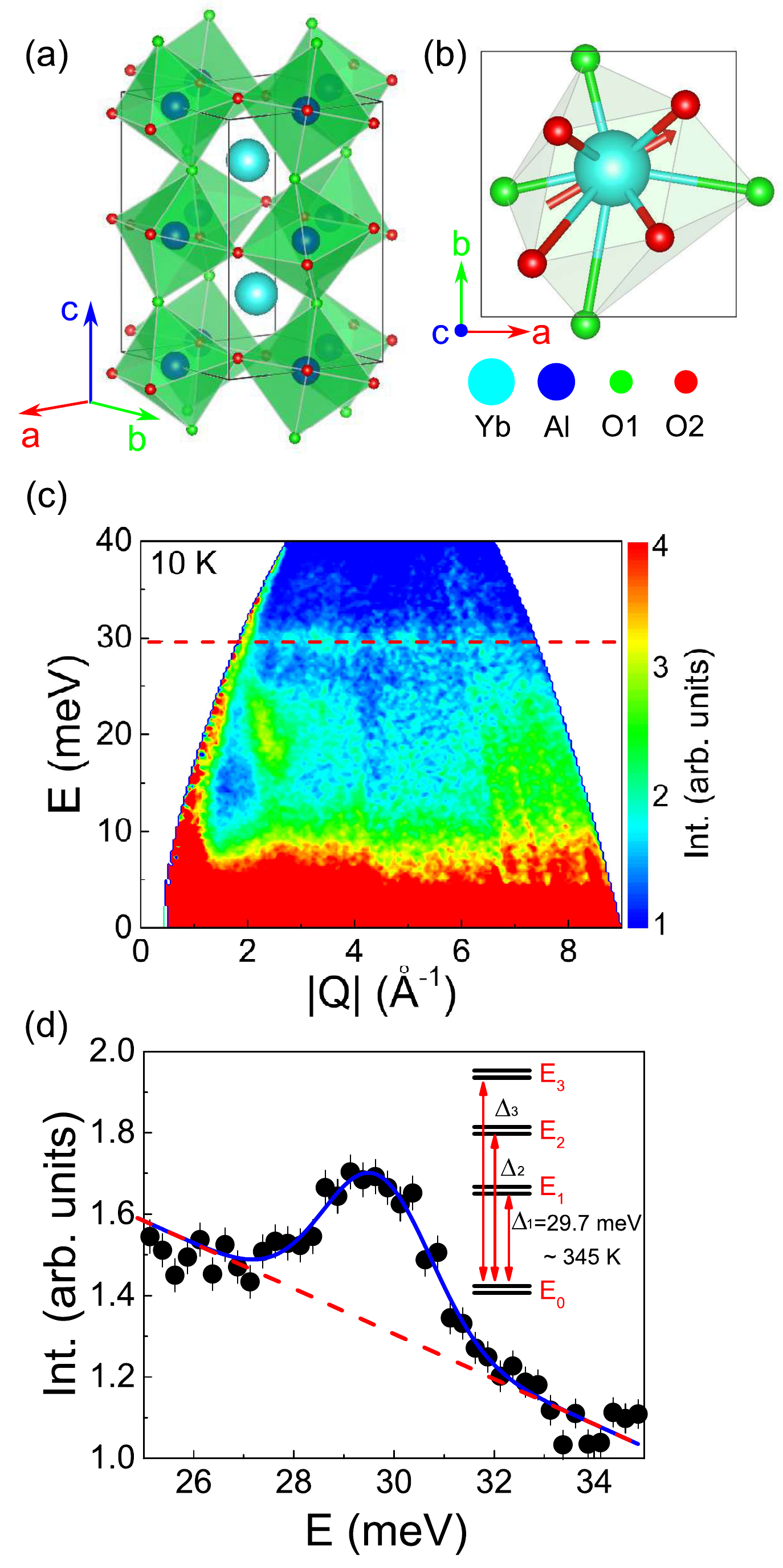}\\
  \caption{(a) Crystal structure of \YbAl, where Yb atoms are surrounded by eight nearby distorted Al-O octahedra. (b) Local chemical environment of Yb$^{3+}$ located at $z = c/4$, considering twelve nearest Oxygen neighbors: four O1 (green circles) sites in the same $z = c/4$ plane, four O2 (red circles) sites above and below the $z = c/4$ plane. The red vector indicates the magnetic moment of Yb$^{3+}$, which lies in the $ab$-plane, with a tilting angle from the $a$-axis. (c) Contour plot of the inelastic neutron scattering spectrum of \YbAl\ measured at 10~K. A flat CEF mode is observed, indicated by the red dashed line. (d) Energy dependent intensity integrated over the wave vector range $|Q|=[3,7]$~{\AA}$^{-1}$. A Gaussian function (blue line) is fit to the data, with a linear background (red dash line) subtracted. Inset: Sketch of the four isolated CEF doublet states of Yb$^{3+}$, where the eight fold degeneracy of $J = 7/2$ ($2J+1=8$) is lifted to four doublet states $E_0$, $E_1$, $E_2$, $E_3$, due to the low point symmetry. The first excited levels are well separated from the ground doublet by $29.7\pm{0.07}$ meV, which is about 345~K.}
  \label{CEF}
\end{figure}
\YbAl\ crystallizes in an orthorhombic distorted perovskite structure, with lattice constants (at room temperature) $a = 5.126$~{\AA}, $b = 5.331$~{\AA}, and $c = 7.313$~{\AA}, using the conventional $Pbnm$ notation~\cite{Buryy2010,Noginov2001}.
As illustrated in Fig.~\ref{CEF}a, the Ytterbium (Yb) ions are surrounded by eight nearby distorted Aluminum-Oxygen (Al-O) octahedra.
Due to the distortion, the point symmetry of the Yb site is lowered from $O_{\rm{h}}$ ($m\bar{3}m$) in the perfect cubic perovskite structure (space group $Pm\bar{3}m$) to $C_{\rm{s}}$ (m).
Therefore, the rare-earth moments ($R$) are constrained either along the $c$-axis, as in the case for $R$ = Er, Tm$^{3+}$~\cite{Deng2015, ke2016anisotropic}, or to the $ab$-plane, as for $R$ = Dy$^{3+}$, Tb$^{3+}$, Yb$^{3+}$~\cite{Tokunaga,Schuchert,Hufner1968,Wu2017}.
No high-symmetry directions within the $ab$-plane are required by the point group symmetry, and in $RM$O$_3$ materials the rare-earth moments are generally aligned to directions that are tilted from the principal $a$- and $b$-axes.
The tilting angle depends on the relative distortion of the eight Al-O octahedra comprising twelve nearest oxygen  neighbors around the Yb$^{3+}$ ion, which also determines the rare-earth CEF splitting.

The calculation of the ground state wave functions and the CEF configuration was based on the point charge model~\cite{McPhase,Stevens1952,Hutchings1964,Rotter2011}.
As in the case of the iso-structural compound \DySc~\cite{Wu2017}, the first twelve nearest oxygen neighbors around the Yb$^{3+}$ ion are considered (Fig.~\ref{CEF}(b)), which keeps the correct local point group symmetry ($C_{\rm{s}}$) of the Yb site.
In this local chemical environment, the eight-fold degenerate $J = 7/2$ ($L = 3$, $S = 1/2$) multiplet ($2J + 1 = 8$) of Yb$^{3+}$ is split into four doublet states.
The four CEF doublet states are best diagonalized when the local Ising axes are chosen along a direction at $\varphi = 22^{\circ}$ titled from the $a$-axis, as indicated by the red vector in Fig.~\ref{CEF}(b).
The calculated ground state wave functions are:
\begin{multline}
\label{CEF_wave_function}
E_{0\pm} =+0.78 |\pm 7/2\rangle-0.03 |\mp 5/2\rangle-0.53 |\pm 3/2\rangle \\ \mp0.14 |\mp 3/2\rangle \pm0.18 |\pm 1/2\rangle+0.23 |\mp 1/2\rangle,
\end{multline}
with the excited levels $E_{1}$, $E_{2}$ and $E_{3}$ separated from the ground doublet at energies $\Delta_{1} = 6.7$~meV, $\Delta_{2} = 24.1$~meV, and $\Delta_{3} = 47$~meV.
We should emphasize here, that although the point charge CEF calculation based on twelve nearest neighbors is only an approximation, and the energy scheme cannot quantitatively reproduce the real ground state wave function, it does qualitatively confirm two important details: i) there is a well separated ground state doublet, which is Ising-like with most contribution from the wave function $|\pm{7/2}\rangle$; ii) the local Ising axis is tilted from the $a$-axis by an angle $\pm\varphi$ of $\sim{22}^{\circ}$.
Both of these are further confirmed by the neutron scattering and magnetization measurements discussed below.

Single crystal inelastic neutron scattering of \YbAl\ was performed, and the spectrum measured at 10~K shows a flat wave vector independent mode (red dashed line in Fig.~\ref{CEF}c), indicating the excitation from the ground state to the first excited CEF level.
The energy dependent intensity integrated over the wave vector range $|Q|=[3,7]$~{\AA}$^{-1}$ is plotted in Fig.~\ref{CEF}d.
A Gaussian function (blue line) is fit to the data, with a linear background (red dash line) subtracted.
The peak is found to be at $29.7\pm{0.1}$~meV, which is about $\sim{345}$~K (inset of Fig.~\ref{CEF}d).
This energy scale is much larger than the CEF calculation above indicates, and suggests well separated ground doublets, which dominate the low temperature magnetic properties.

\subsection{Magnetization}

\begin{figure}
  \centering
  \includegraphics[width=1.0\columnwidth]{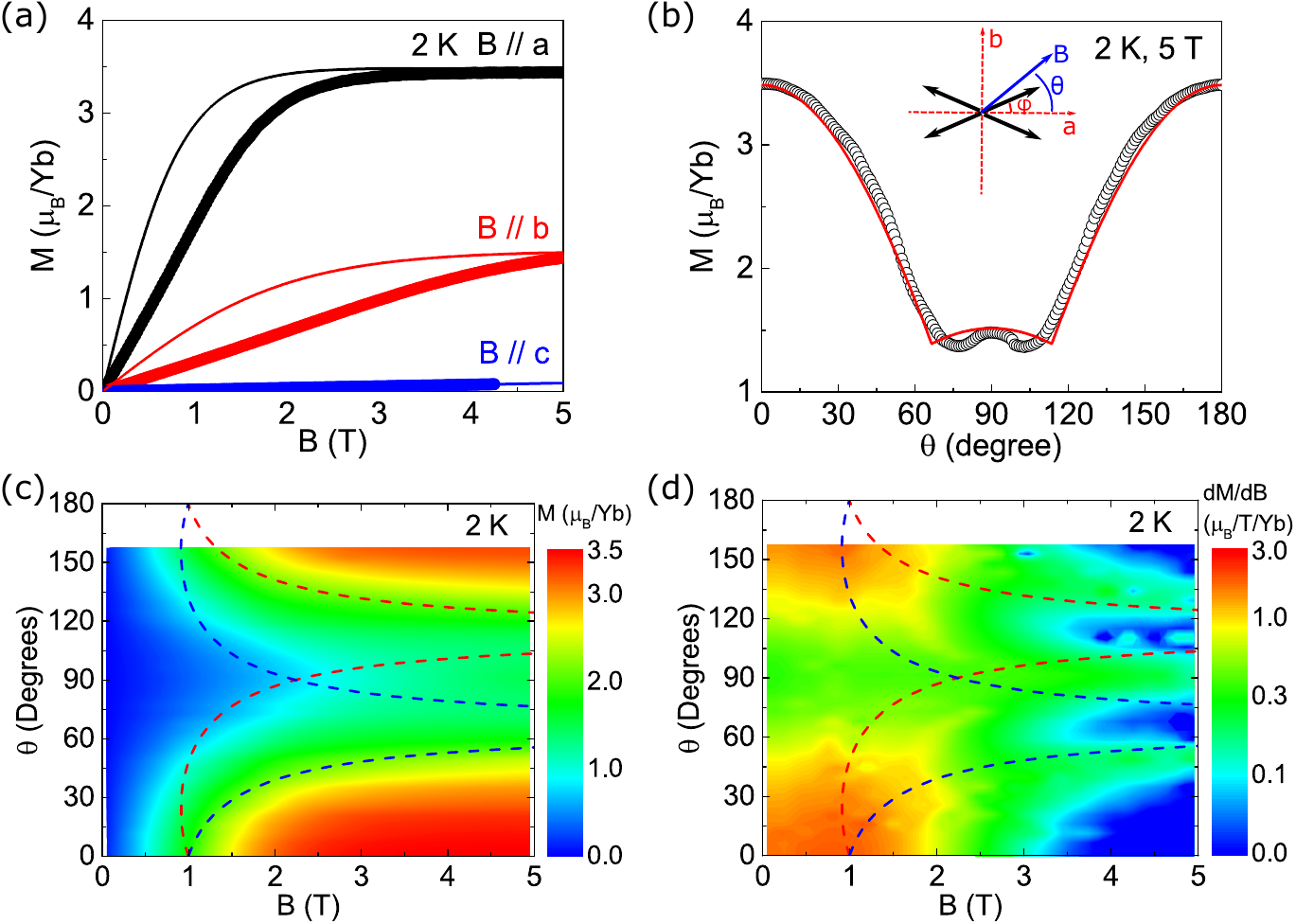}\\
  \caption{(a) Field dependent magnetization $M$ of \YbAl, measured at 2~K, with the field along different principal axes.
  The solid lines are the calculated Brillouin functions, as explained in the text.
  (b) Angle-dependent magnetization measured at $T = 2$~K and $B = 5$~T. The red line represents the fit, as explained in the text. Inset: Schematic view of the Yb magnetic moment configuration in the ordered state, where $\varphi$ is the angle between the Ising moments and the $a$-axis, and the angle $\theta$ indicates the direction of the applied field in the $ab$-plane.
  (c)-(d) Contour plot of the measured magnetization $M$ and magnetic susceptibility $dM/dB$ at 2~K, with magnetic field rotating in the $ab$-plane.
  The red and blue dashed lines indicate the angle dependent critical fields.}
  \label{MB}
\end{figure}

The field dependent magnetization $M$ of \YbAl\  measured at temperature $T = 2$~K is shown in Fig.~\ref{MB}(a).
With the field applied along different principal crystal axes, significant anisotropy is observed between the $ab$-plane, and the $c$-axis. The high field saturation moments along the $a$ and $b$ directions are more than one order larger than the moment along $c$-axis, confirming that the Yb$^{3+}$ magnetic moments are lying in the $ab$-plane.
A further measurement of the anisotropy in the $ab$-plane was performed with the horizontal rotator.
As presented in Fig.~\ref{MB}(b), the angle dependence of the magnetization measured in magnetic field $B = 5$~T has two minima at $\theta = 90\pm\varphi$, where $\varphi$ is the angle between the Ising moments and the $a$-axis, and the angle $\theta$ indicated the direction of the applied field in the $ab$-plane. The angular dependence can be described as:
\begin{align}
\label{MvsB}
M&=\dfrac{M_{\rm{s}}}{2}(\lvert\cos(\varphi-\theta)\rvert+\lvert\cos(\varphi+\theta)\rvert) ,
\end{align}
where $M_{\rm{s}}$ is the saturation moment.
The result of the fitting is shown as the red line in Fig.~\ref{MB}b.
With $M_{\rm{s}}=3.8\mu_{\rm{B}}\rm{/Yb}$, and $\varphi=23.5^{\circ}$, the calculated curve matches well the experimental magnetization.
Since the experimental temperature ($T=2$~K) and magnetic field ($B < 5$~T) are much smaller than the energy scale of the first excited CEF level $\Delta_{1} = 29.7$~meV, the isolated ground state doublet can be described as an effective spin $S = 1/2$,
where $M = g_{\rm{eff}}\mu_{\rm{B}}\cdot{S}$.
The calculated Brillouin functions are shown in Fig.~\ref{MB}a (solid lines), and the saturation moments for different field directions are:
\begin{align}\label{BF}
M & _{a}= M_{\rm{s}}\cos\varphi\backsimeq3.47 \mu_{\rm{B}}\rm{/Yb},\\
M & _{b}= M_{\rm{s}}\sin\varphi\backsimeq1.54 \mu_{\rm{B}}\rm{/Yb},\\
M & _{c}\backsimeq 0.23 \mu_{\rm{B}}\rm{/Yb}.
\end{align}
This could be equivalently expressed with anisotropic $g$-factors as:
\begin{align}
\label{geff}
M _{\rm{z}}=M_{\rm{s}}= &3.8 \mu_{\rm{B}}\rm{/Yb},&\ \ &g^z_{\rm{eff}}=7.6,\\
M _{\rm{\bot}}=M_{\rm{c}}= &0.23 \mu_{\rm{B}}\rm{/Yb},&\ \ &g^{\bot}_{\rm{eff}}=0.46,
\end{align}
where $z$ is chosen along the local moment Ising axis, and $xy$ are the perpendicular directions.
At low magnetic field, the measured magnetization curve is lower than the calculated Brillouin function, and this mismatch suggests the existence of an additional AF magnetic correlation in the system, which is missing in the calculation.
With increasing field, all magnetic moments will eventually align with the field, and the magnetization curve and the calculated Brillouin function would overlap. This saturation field for polarization is also angle dependent, which can be expressed as:
\begin{align}
\label{crit_field8}
B_{\rm{s}}^{a}&=B_{\rm{s}}/\cos\varphi,\\
\label{crit_field9}
B_{\rm{s}}^{b}&=B_{\rm{s}}/\sin\varphi.
\end{align}
Here, $B_{\rm{s}}^{a}$ and $B_{\rm{s}}^{b}$ are the polarization fields needed for magnetic field applied along the $a$- and $b$-axes, respectively, and $B_{\rm{s}}$ is the polarization field when the applied magnetic field is along the local Ising axis.
Using the above relations~(\ref{crit_field8}) and (\ref{crit_field9}), and knowing $\varphi=23.5^{\circ}$, one can calculate that $B_{\rm{s}}\simeq0.9$~T.
The whole field and angle dependence of the magnetization and magnetic susceptibility are shown in Fig.~\ref{MB}(d) and (d).
The red and blue dashed lines are the calculated critical fields of relations~(\ref{crit_field8}) and (\ref{crit_field9}), and  one can see that the overall field and angle dependence of the magnetization is well reproduced.

Compared to the CEF calculation discussed earlier, the experimental magnetization measurement suggests a ground state wave function with a much stronger Ising anisotropy.
Such wave functions can be estimated more quantitatively based on the experimental saturation moment, extracted from above relation~(\ref{geff}).
A generic ground state wave function can be expressed as:
\begin{multline}
\label{ground_states}
E_{0\pm} =\alpha |\pm 7/2\rangle+\beta |\pm 5/2\rangle+\gamma |\pm 3/2\rangle+ \delta |\pm 1/2\rangle\\+\alpha' |\mp 7/2\rangle+\beta' |\mp 5/2\rangle+\gamma'|\mp 3/2\rangle+ \delta' |\mp 1/2\rangle.
\end{multline}
The point group symmetry further imposes the constraints $\beta=\alpha'=0$.
Based on this, the saturation moment along the Ising axis ($z$) is expressed as:
\begin{multline}
M_{z}=g\langle E_{0\pm}|J_{\rm z}|E_{0\pm}\rangle \\=
\frac{8}{7}\cdot\left[\frac{7}{2}\alpha^2-\frac{5}{2}\beta'^2+\frac{3}{2}\left(\gamma^2-\gamma'^2\right)  + \frac{1}{2}\left(\delta^2-\delta'^2\right)\right],
\end{multline}
where $g=8/7$ is the the Land{\'e} factor of Yb$^{3+}$.
The upper and lower limits of the contributions from states $|\pm 7/2\rangle$ can be estimated, considering the two extreme cases.
For the upper limit value of $\alpha$, assuming that all contributions from states $|\pm 3/2\rangle$ and $|\pm 1/2\rangle$ are zero ($\gamma=\gamma'=0$, and $\delta=\delta'=0$), then to make $M_{z}=3.8 \mu_{\rm B}\rm/Yb$, one has $\alpha=0.985$, $\beta'=0.171$.
Since the CEF configurations are determined by the nearby charges, the Yb ground state wave function and moment configuration should be applicable to other iso-structural perovskites as well, such as in \YbFe, where Fe$^{3+}$ provides an almost identical charge environment as Al$^{3+}$ here.

\subsection{Magnetic dipole-dipole interaction}

\begin{figure}
  \centering
  \includegraphics[width=0.6\columnwidth]{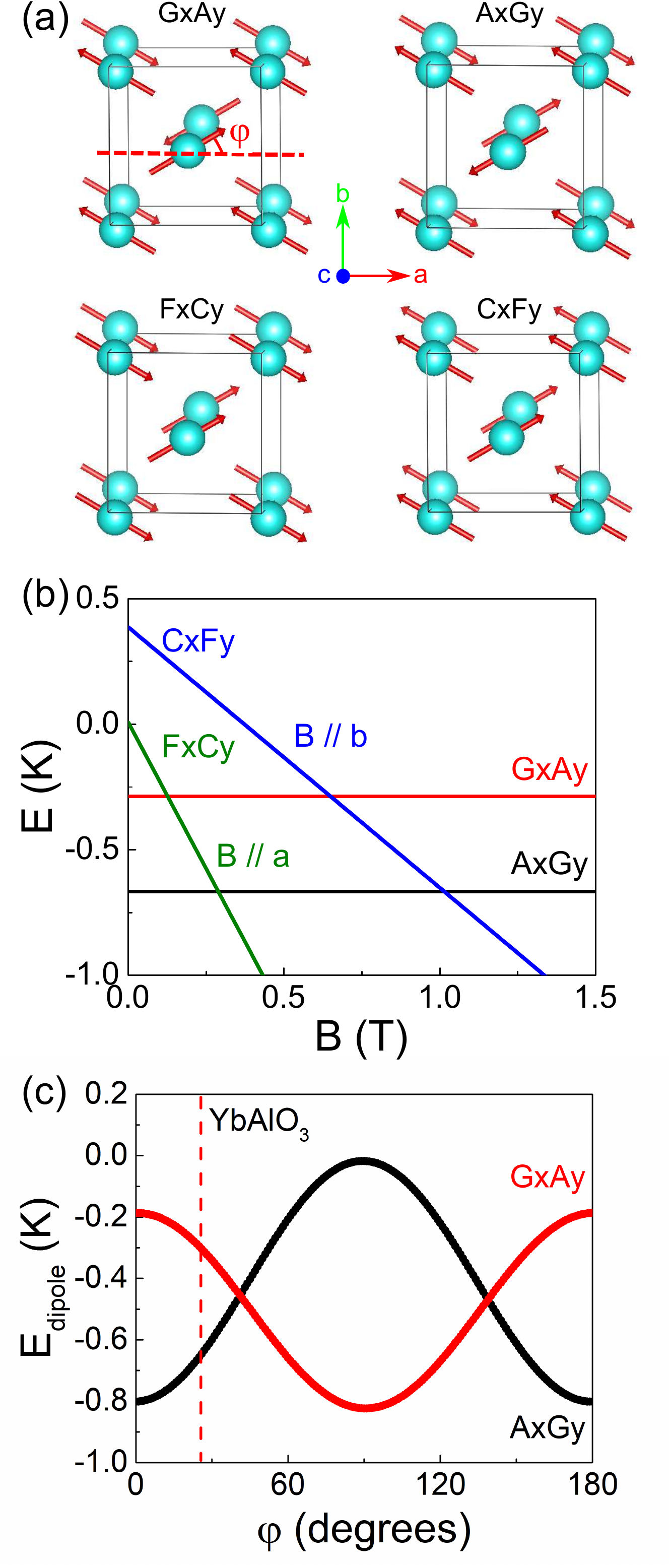}\\
  \caption{(a) Configurations of the four symmetry allowed magnetic structures.
  The red vectors indicate the Yb magnetic moment directions, which are tilted by angle $\varphi$ from the $a$-axis.
  (b) Calculated field dependence of the dipole-dipole energy for each of the four magnetic structures. For magnetic field applied along the $a$- and $b$-axes, the ferromagnetic $FxCy$ and $CxFy$ configurations will be the new ground states, with critical fields of 0.3~T and 1.0~T, respectively.
  (c) Calculated angle dependence of the dipole-dipole energy for the two antiferromagnetic structures, $GxAy$ and $AxGy$.
  With $\varphi = 23.5^{\circ}$, the configuration $AxGy$ is selected as the ground state for \YbAl, as indicated by the red dashed line.
  }
  \label{dipoleFig}
\end{figure}

Within the given crystal structure, four different magnetic structures with propagation vector $k = 0$ ($AxGy$, $GxAy$, $FxCy$ and $CxFy$) are allowed through representation analysis, as shown in Fig.~\ref{dipoleFig}(a).
It was reported that the configuration $AxGy$ is chosen in zero field below
the N{\'e}el temperature $T_{\rm{N}} = 0.88$~K~\cite{Radhakrishna1981}.
Due to the large local saturation moment of Yb$^{3+}$ at low temperatures, $M_{\rm{s}}\simeq3.8\mu_{\rm{B}}{\rm{/Yb}}$, the dipole-dipole interaction is not negligible.
For each of the four magnetic structures ($GxAy$, $AxGy$, $FxCy$ and $CxFy$), the dipole-dipole energy
\begin{equation}
\label{dipoleFormulaEq}
E_{\rm{dip}}=-\frac{\mu_{0}}{4\pi}\sum_{\rm{i}}\frac{1}{|\vec{\mathbf{r}}_{\rm{i}}|^3}\cdot \left[3\left(\vec{\mathbf{m}}_{0}\cdot\hat{\mathbf{r}}_{\rm{i}}\right)\left(\vec{\mathbf{m}}_{\rm{i}}\cdot \hat{\mathbf{r}}_{\rm{i}}\right) - \left(\vec{\mathbf{m}}_{0}\cdot\vec{\mathbf{m}}_{\rm{i}}\right)\right]
\end{equation}
was calculated, where $\mu_{0}$ is the vacuum permeability, and $\hat{\mathbf{r}}_{\rm{i}}=\vec{\mathbf{r}_{\rm{i}}}/|\vec{\mathbf{r}}_{\rm{i}}|$.
Here, 10 near neighbor moments $\vec{\mathbf{m}}_{\rm{i}}$ (eight within distance $\sim5.7$~\AA\ in the $ab$-plane, and two near neighbors of distance $\sim4$~\AA\ along the $c$-axis) around the moments $\vec{\mathbf{m}}_0$ centered at the origin were included in the sum.
It was found that adding more terms to the sum (for neighbors at larger distances) only resulted in negligibly small variations of the dipole energy.
In zero field, the calculated dipole energies for different configurations are (Fig.~\ref{dipoleFig}(b)): -0.67~K, -0.29~K, -0.01~K, and 0.38~K, respectively.
It is clear that the configuration $AxGy$ is selected as the ground state at low enough temperatures, and the AF magnetic ordering temperature $T_{\rm{N}} = 0.88$~K is consistent with the energy gain ($E_{\rm{dip}}(AxGy) = -0.67$ K) estimated from the dipole-dipole interaction.

When a magnetic field is applied along the $a$ or $b$ directions, the configurations $FxCy$ and $CxFy$ are favored by the Zeeman interaction~\cite{Wu2017}, making them the new ground states.
The calculated critical fields are $B_{\rm{dip}}^{\rm{a}}\sim{0.3}$~T and $B_{\rm{dip}}^{\rm{b}}\sim{1.0}$~T, respectively.
Similar calculations were performed for \DySc\ as well, and a good agreement between the calculation and the experiments was observed~\cite{Wu2017}.
However, for \YbAl, these calculated critical fields only account for about $30\%-50\%$ of the the polarization fields extracted from the field dependent magnetization measurements, where $B_{\rm{s}}^{\rm{a}}=1.1$~T, and $B_{\rm{s}}^{\rm{b}}=2.2$~T.
Noticing that the energy differences between configurations $AxGy$ and $FxCy$ (or between $GxAy$ and $CxFy$) are from the interactions in the plane (or interactions along the $c$-axis), this discrepancy suggests that an additional exchange coupling exists along the $c$-axis in \YbAl. Further, this indicates there may exist a quantum critical region between $0.3\sim1.1$~T for $B\parallel{a}$ (or $1.0\sim2.2$~T for $B\parallel{b}$), where the static  magnetic order ($AxGy$) in the $ab$-plane is suppressed~\cite{Wu2019}.

\begin{figure}
  \centering
  \includegraphics[width=0.8\columnwidth]{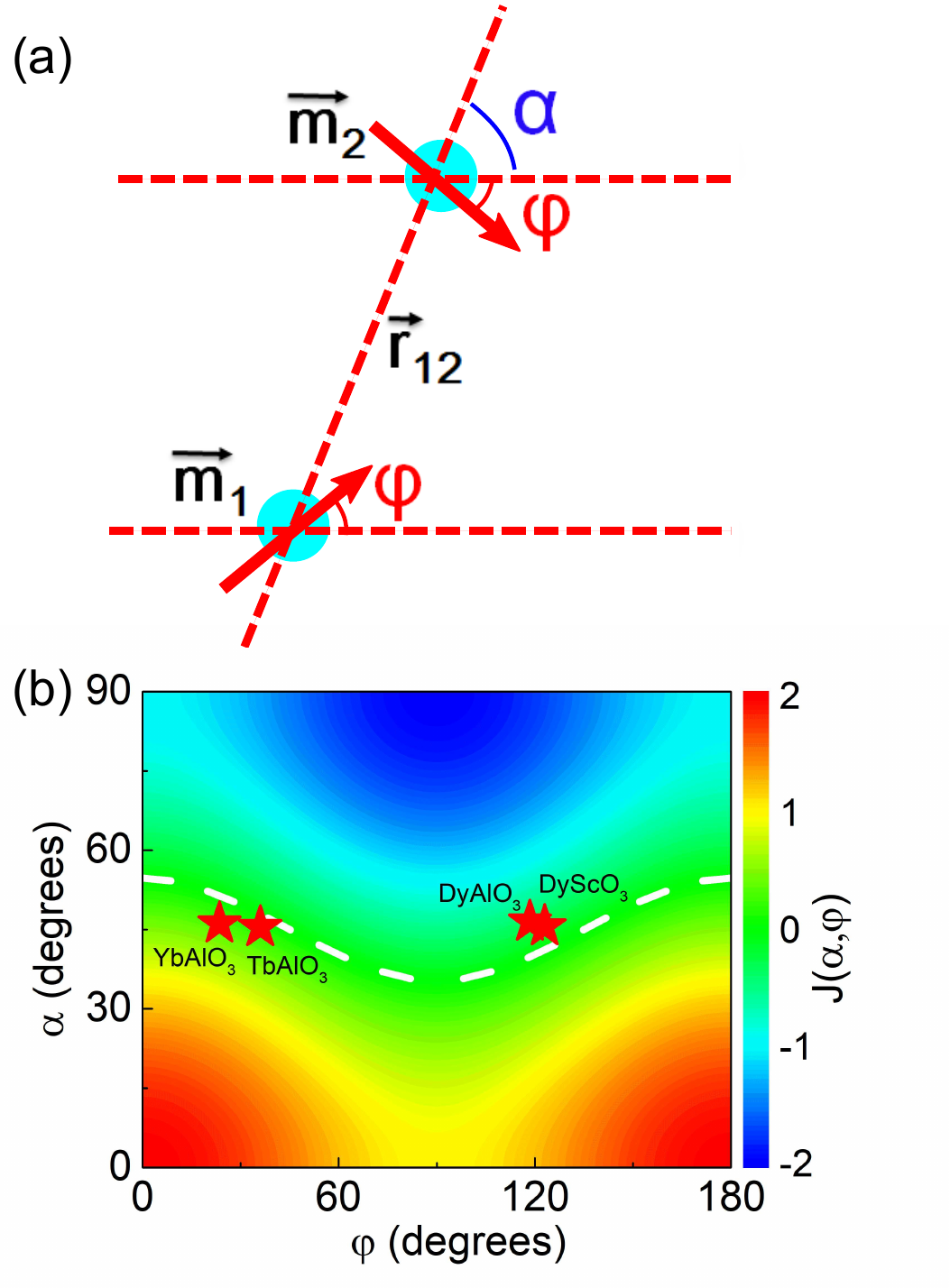}\\
  \caption{(a) Sketch of the magnetic dipole-dipole interaction between two Ising magnetic moments $\vec{\mathbf{m}}_1$ and $\vec{\mathbf{m}}_2$.
  The two moments are separated by distance $\vec{\mathbf{r}}_{12}$, and make an angle $\pm\varphi$ with the $a$-axis.
  The angle $\alpha=47.73^{\circ}$ is the angle between vector $\vec{\mathbf{r}}_{12}$ and the crystal $a$-axis.
  (b) Contour plot of the calculated dipole-dipole interaction as a function of both angles $\alpha$ and $\varphi$.
  The white dashed line indicates the line of `magic' angles, where the dipolar interaction vanishes between the two nearby Ising moments.
  The stars indicate that the rare-earth orthorhombic perovskites of interest here are all located in the vicinity of these `magic' angles, with very weak dipolar interaction in the $ab$-plane.}
  \label{DipoleContourFig}
\end{figure}

It may also be noticed that the two configurations $GxAy$, $AxGy$ should be degenerate if there are no interactions in the $ab$-plane.
Therefore, it is the intra-plane interaction that lifts this degeneracy, and finally selects the static magnetic ordering pattern in the $ab$-plane.
This intra-plane dipolar interaction depends on the relative tilting angle $\varphi$ of the Ising moments~\cite{Kappatsch}, and depending on the value of $\varphi$, the ground state may be the configuration $GxAy$ or $AxGy$.
For the case of \YbAl\ with $\varphi=23.5^{\circ}$, the configuration $AxGy$ is selected, while in the iso-structural compound \DySc\ with $\varphi=90^{\circ}+28^{\circ}=118^{\circ}$, configuration $GxAy$ is selected~\cite{Wu2017}.

Similar calculations may also be applied to other rare-earth based perovskites, where the dipole-dipole interaction plays an important role at low temperatures.
Shown in Fig.~\ref{DipoleContourFig}(a) is the magnetic dipole-dipole interaction between two Yb Ising magnetic moments $|\vec{\mathbf{m}}_1|=|\vec{\mathbf{m}}_2|=|\vec{\mathbf{m}}|$, separated by distance $|\vec{\mathbf{r}}_{12}|=|\vec{\mathbf{r}}|$.
Within the given lattice symmetry, the Ising moments are only allowed to tilt toward each other by the same angle $\pm\varphi$ from the $a$-axis, while $\alpha=47.73$ is the angle between vector $\vec{r}_{12}$ and the $a$-axis.
This simplifies the equation for the dipole interaction~(\ref{dipoleFormulaEq}) to:
\begin{equation}
\label{dipoleInteraction}
E_{\rm{dip}}=-\frac{\mu_{\rm{0}}}{4\pi}\frac{1}{|\vec{\mathbf{r}}|^{3}}|\vec{\mathbf{m}}|^{2}J(\alpha,\varphi),
\end{equation}
where
\begin{equation}
J(\alpha,\varphi)=\cos^{2}\varphi+3\cos^{2}\alpha-2.
\end{equation}
For $\varphi=0$, there is a `magic' angle of $\alpha=54.7^{\circ}$, at which point the dipolar interaction vanishes.
By adding an additional degree of freedom (angle $\varphi$), the `magic' angle becomes a `magic' line $\alpha(\varphi)$.
Shown in Fig.~\ref{DipoleContourFig}(b) is the contour plot of the calculated values of $J(\alpha,\varphi)$ as functions of $\varphi$ and $\alpha$.
The dipole interaction vanishes everywhere on the white dashed line, where
\begin{equation}
  \label{dipoleEq}
  J(\alpha,\varphi)=0.
\end{equation}

It turns out that \YbAl, as well as a few other materials for which the angles $\alpha$ and $\varphi$ were reported~\cite{Tokunaga,Schuchert,Hufner1968,Wu2017}, are located in the vicinity of this `magic' line,
as indicated by the stars in Fig.~\ref{DipoleContourFig}(b).
More other iso-structural rare-earth perovskite systems could be fit to this diagram, suggesting a whole new family of rare-earth based one-dimensional magnets.

\subsection{Magnetic diffuse scattering}

\begin{figure}
  \centering
  \includegraphics[width=1\columnwidth]{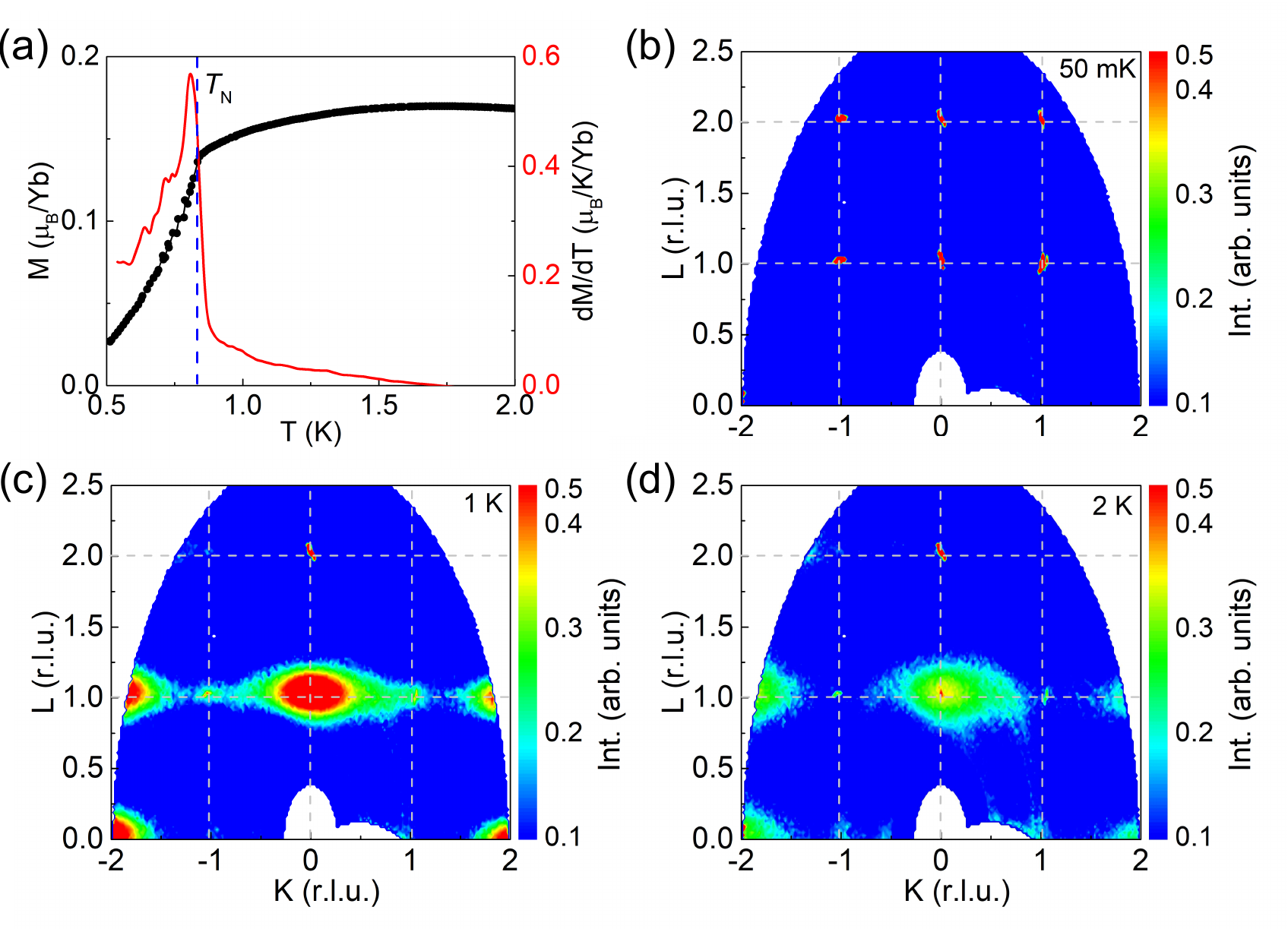}\\
  \caption{(a) Magnetization $M$ (black) and temperature derivative $dM/dT$ (red) of \YbAl, with applied field $B = 0.1$~T along the $a$-axis.
  The blue dashed line indicates the antiferromagnetic transition at $T_{\rm{N}} = 0.85$~K.
  (b)-(d): Contour plot of the magnetic scattering of \YbAl\ in the $(0KL)$ plane, integrated over wave vector $H = [-0.2,0.2]$ r. l. u., and energy $E = [-0.1,0.1]$~meV, at different temperatures, 50~mK (b), 1.0~K (c) and 2.0~K (d), respectively.}
  \label{Diffuse}
\end{figure}

\begin{figure}
  \centering
  \includegraphics[width=0.6\columnwidth]{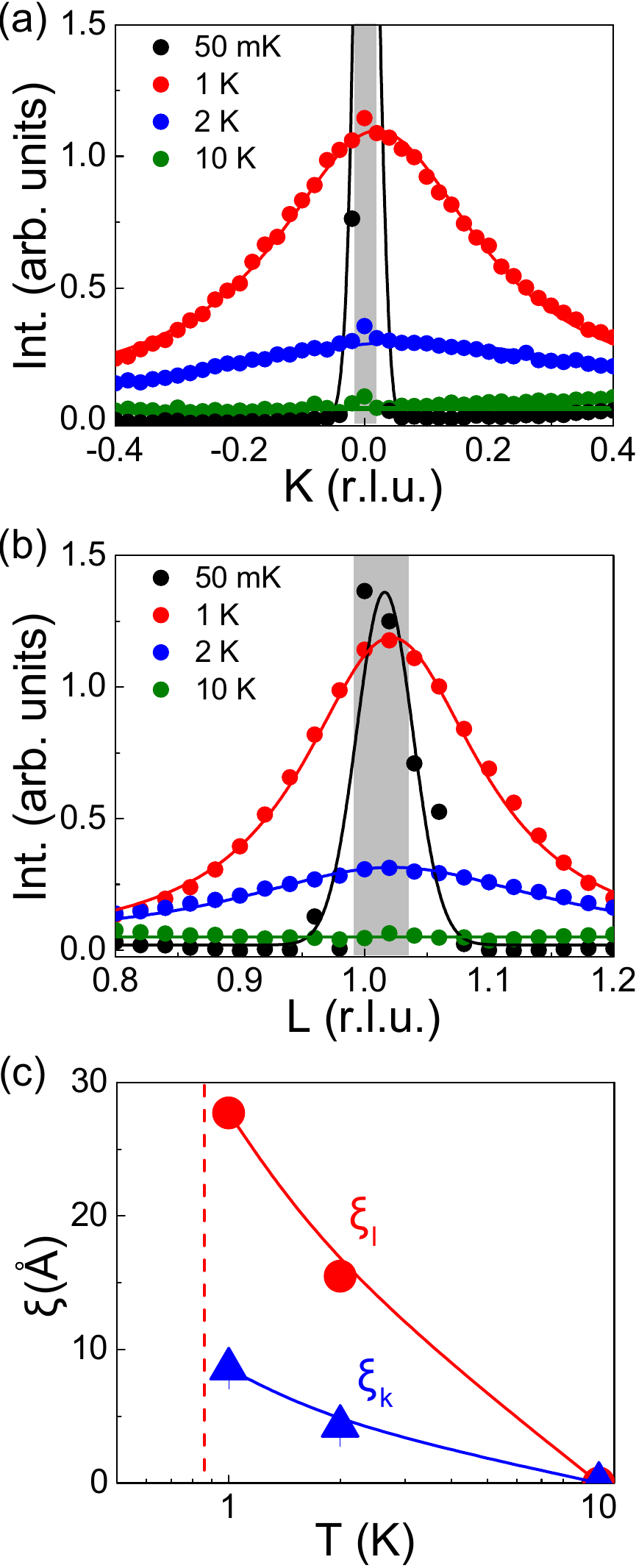}
  \caption{(a) Constant energy cut along wave vector $K$, integrated over $E = [-0.1,0.1]$~meV, $H = [-0.2,0.2]$ r. l. u., and $L = [0.9,1.1]$ r. l. u.
  (b) Constant energy cut along wave vector $L$, integrated over $E = [-0.1,0.1]$~meV, $H = [-0.2,0.2]$, and $K = [-0.2,0.2]$ r. l. u.
  The instrumental resolution (gray bar) is estimated from the full width at the half maximum (FWHM) of the magnetic (001) peak measured at 50~mK.
  The solid lines are the fit with Lorentzian functions, as explained in the text.
  (d) The temperature dependent correlation length along different directions.
  The red dashed line indicates the magnetic transition at 0.88~K.}
  \label{Correlation}
\end{figure}

The temperature dependent magnetization $M$ of \YbAl\ was measured down to 0.5~K in an applied field $B=0.1$~T along the $a$-axis. These data, and the derivative $dM/dT$, are shown in Fig.~\ref{Diffuse}(a). An abrupt decrease in the magnetization is observed with lowering the temperature, and the antiferromagnetic transition temperature is determined through the peak in the curve of $dM/dT$, with $T_{\rm{N}} \simeq {0.88}$~K (blue dashed line in Fig.~\ref{Diffuse}(a)), which is consistent with a previous report~\cite{Radhakrishna1981,Wu2019}. Single crystal neutron scattering was performed in the $(0KL)$ scattering plane, at different temperatures below and above the magnetic transition, as presented in Fig.~\ref{Diffuse}(b)-(d).
For temperatures below the AFM transition, magnetic peaks are well established, and the peak width is limited by the instrumental resolution (Fig.~\ref{Diffuse}(b)).
However, above the phase transition, broad diffuse scattering is observed near magnetic wave vectors such as $\mathbf{Q} = (0,0,1)$, which is much broader than the nuclear diffraction peaks such as $\mathbf{Q} = (0,0,2)$ (Fig.~\ref{Diffuse}(c) and (d)).
Further, this broad magnetic diffuse scattering shows a clear ellipse like shape with long axis along wave vector $(0K0)$, suggesting a very anisotropic correlation in the $(0KL)$ scattering plane.
For a more quantitative analysis, constant energy cuts along wave vectors $K$ and $L$ at different temperatures are presented in Fig.~\ref{Correlation}(a) and (b).
The instrumental resolution (gray bar) is estimated from the full width at half maximum (FWHM) of the magnetic (001) peak measured at 50~mK.
The following function~\cite{Zaliznyak2015,Wu2017} was adopted to describe the magnetic structure factor:
\begin{equation}
\label{lattice_lorentzian}
S(Q)\propto\frac{\rm sinh(c/\xi_{l})}{\rm cosh(c/\xi_{l})-\cos(\pi(l-1))} \cdot
\frac{\rm sinh(b/\xi_{k})}{\rm cosh(b/\xi_{k})-\cos(\pi{k})},
\end{equation}
where $\xi_{k}$ and $\xi_{l}$ are the correlation lengths in real space along the $b$- and $c$-axes.
Since the variation of the diffuse scattering with the wave vector $\mathbf{Q}$ is much stronger than that of the magnetic form factor, we have ignored the latter for the fits shown in Fig.~\ref{Correlation}.
The fitted correlation lengths at different temperatures are shown in Fig.~\ref{Correlation}(c).
As expected, both correlation lengths $\xi_{k}$ and $\xi_{l}$ increase while approaching the magnetic ordering temperature from above.
However, the building up of the correlation along the $c$-axis is much faster than along the $b$-axis.
This significant anisotropy is as expected from the estimate of the dipole-dipole interaction discussed in the previous section, and it clearly reflects the one-dimensional character of \YbAl.

\section{Discussion and Conclusion}

In summary, magnetic properties of \YbAl\ have been studied through a combination of CEF calculations and measurements of magnetization and single crystal neutron elastic scattering.
All our results are consistent with a well separated Ising Yb$^{3+}$ ground state doublet, whose wave function mostly consists of $|\pm7/2\rangle$ states.
The local easy axes make an angle $\pm\varphi$ with the $a$-axis, as long as the temperature scale is smaller than the first excited CEF level of 29.7~meV~$\sim{345}$~K.
The Yb$^{3+}$ Ising moments order magnetically below $T_{\rm{N}} = 0.88$~K, and the AF magnetic ground state $AxGy$ is stabilized by the dipole-dipole interaction.
Since the magnetic ground state resulting from the CEF is controlled by the near neighbor point charge configuration, it is very little affected by substitution of Al$^{3+}$ by other ions such as Fe$^{3+}$, as in iso-structural compound \YbFe.
Thus, it is very likely that, in \YbFe, Yb moments share similar Ising ground states with their local easy axis in the $ab$-plane, which is very different from the scenario proposed in~\cite{brown1993Yb} where Yb moments are rotating between the $c$-axis and the $ab$-plane.

Further analysis of the dipolar interaction suggests that the series of iso-structural rare-earth perovskites composes a new family of one dimensional magnets.
This observation is further supported by the magnetic diffuse scattering observed at low temperature, where critical fluctuations indicate mostly a one-dimensional correlation along the $c$-axis.

\YbAl\ shares many similarities with the compound \DySc\ which had been studied earlier~\cite{Wu2017}.
In both systems, the ground states show a significant Ising like anisotropy with
$M_{\|}/M_{\bot}=\langle E_{0\pm}|J_{\rm z}|E_{0\pm}\rangle/\langle E_{0\pm}|J_{\rm xy}|E_{0\pm}\rangle\gg1$. Similarly, a direct corollary is that the magnetic fluctuations are dominated by their longitudinal component, which is about two orders larger than the transverse fluctuations
($M_{\|}/M_{\bot}\simeq{273}$ for \YbAl\ and $M_{\|}/M_{\bot}\simeq{400}$ for \DySc).
Thus, any transverse fluctuations such as spin waves (or magnons), which are usually observed in Heisenberg spin systems~\cite{Mourigal2013}, will be negligible in both \YbAl\ and \DySc.

However, whereas longitudinal fluctuations are forbidden in \DySc\ by a selection rule~\cite{Wu2017}, they are being observed in \YbAl~\cite{Wu2019}.
The ground state wave functions in \YbAl\ (Eqs.~(\ref{CEF_wave_function}) and (\ref{ground_states})) ensure that the matrix elements connecting the states of moments `up' and `down' are non-zero, that is,
$\langle E_{0\mp}|S^{+}, S^{-}| E_{0\pm}\rangle=\alpha\beta'+...\neq{0}$.
This crucial point allows spin flip exchange terms in the low energy theory of the ground state doublets.
Thus, distinct from \DySc, longitudinal spinon excitations are visible to neutrons in \YbAl, making it a novel promising quantum magnet for further exploring the low-dimensional critical dynamics.

\begin{acknowledgments}
We would like to thank G. Loutts for providing the single crystal \YbAl\ sample, and Z. Wang, C. D. Batista, I. Zaliznyak, A. Tsvelik, F. Ronning, M. C. Aronson, E-J. Guo and J. Sheng for helpful discussions.
This research used resources at the High Flux Isotope Reactor and Spallation Neutron Source, a DOE Office of Science User Facility operated by the Oak Ridge National Laboratory. Research supported in part by the Laboratory Directed Research and Development Program of Oak Ridge National Laboratory, managed by UT-Battelle, LLC, for the U.S. Department of Energy. This work is partly supported by the U.S. Department of Energy (DOE), Office of Science, Basic Energy Sciences (BES), Materials Science and Engineering Division. S.E.N. acknowledges support from the International Max Planck Research School for Chemistry and Physics of Quantum Materials (IMPRS-CPQM). A.D.C. is partially supported by the U.S. Department of Energy, Office of Science, Basic Energy Sciences, Materials Sciences and Engineering Division. L.V. acknowledges Ukrainian Ministry of Education and Sciences for partial support under project ``Feryt''. M.B. would like to thank the DFG for financial support from project BR 4110/1-1.
\end{acknowledgments}

\bibliography{YbAlO3_PRB}

\end{document}